\documentclass[superscriptaddress, reprint, amsmath, amssymb, aps, pra, floatfix]{revtex4-2}
\usepackage{graphicx} % Required for inserting images

\usepackage{xcolor} % for colored text
\usepackage{hyperref} % for hyperlinks
\usepackage{orcidlink} % for ORCID iDs
\hypersetup{
	colorlinks=true,
	linkcolor=blue,
	filecolor=magenta,
	urlcolor=magenta,
}

\definecolor{dgreen}{rgb}{0.0,0.5,0.0}
\definecolor{pink}{rgb}{1,0,0.9}

\newcommand{\change}[1]{{#1}}
\usepackage[normalem]{ulem}

\begin{abstract}
We study the SU(4)-symmetric Heisenberg model on a cubic lattice with spatially anisotropic magnetic couplings. We utilize several approaches based on the tensor-network representation of many-body wave functions, which enable accurate analysis of ground-state properties of the model in different regimes of spatial anisotropy including the fully isotropic three-dimensional case.
Our results point to the persistence of the dimerized color-ordered phase throughout \change{a wide} range of magnetic couplings.
\end{abstract}

\begin{document}

\title{Dimerization in the SU(4) Heisenberg model on the cubic lattice: iPEPS study}

\author{I.V. Lukin\orcidlink{0000-0002-8133-2829}
}
\email{illya.lukin11@gmail.com}
\affiliation{Akhiezer Institute for Theoretical Physics, NSC KIPT, Akademichna 1, 61108 Kharkiv, Ukraine}
\affiliation{Haiqu, Inc., 95 Third Street, San Francisco, California 94103, USA}

\author{A.G. Sotnikov\orcidlink{0000-0002-3632-4790}
}
\email{a\_sotnikov@kipt.kharkov.ua}
\affiliation{Akhiezer Institute for Theoretical Physics, NSC KIPT, Akademichna 1, 61108 Kharkiv, Ukraine}
\date{\today}
\affiliation{Education and Research Institute ``School of Physics and Technology'', Karazin Kharkiv National University, Svobody Square 4, 61022 Kharkiv, Ukraine}

\maketitle

\section{Introduction}

SU($N$)-symmetric quantum many-body systems comprise one of the central fascinating research areas in modern condensed matter physics. They offer rich phase diagrams and exotic quantum phases that extend far beyond conventional SU(2) spin systems. 
The experimental realization of high spin symmetries has been achieved on several platforms, most notably with cold gases of alkaline-earth(-like) atoms, where nuclear spin degrees of freedom can be decoupled from electronic motion, effectively realizing SU($N$) symmetric interactions with $N$ up to 10~\cite{PhysRevLett.98.030401, PhysRevLett.105.030402}, as well as, more recently, with ultracold molecules reaching even higher values of $N$ up to 36~\cite{Mukherjee2025}. Additionally, transition metal compounds with orbital degeneracies naturally exhibit SU(4) symmetry when spin and orbital degrees of freedom are treated on an equal footing, as observed in certain vanadium and titanium oxides~\cite{Kugel1982, PhysRevLett.81.3527, PhysRevLett.121.097201}.
From a theoretical standpoint, SU($N$)-symmetric models provide fundamental insights into quantum criticality, entanglement properties, and the interplay between symmetry and topology in quantum many-body systems~\cite{Caz2014RPP, Ibarra-Garcia-Padilla2025}. The computational study of these models, however, presents significant challenges due to the exponential growth of the Hilbert space dimension with $N$, necessitating the development of advanced numerical techniques that can efficiently capture the entanglement structure of quantum states.

Tensor network methods \cite{orus2019tensor, banuls2023tensor, okunishi2022developments} have shown great promise in the study of strongly correlated quantum systems, including SU($N$)-symmetric models. These theoretical approaches approximate the ground-state wave function or thermal density matrix as a contraction of a large number of small-rank tensors—the tensor network. This network of tensors is chosen in a form that faithfully represents the structure of correlation and entanglement of the ground state. For one-dimensional (1d) quantum lattice systems, the most widely used tensor network ansatze are matrix product states (MPS), which also comprise the basis of the celebrated density matrix renormalization group (DMRG) algorithm~\cite{white1992density, schollwock2011density}. Direct generalization of MPS to two-dimensional (2d) systems in the thermodynamic limit are infinite projected entangled pair states (iPEPS)~\cite{verstraete2008matrix, cirac2021matrix, nishino2001two, verstraete2004renormalization} (for practical introductions, see also Refs.~\cite{orus2014practical, bruognolo2021beginner}). The iPEPS approaches have been successfully employed to study highly correlated 2d problems: the Heisenberg model on the kagome lattice~\cite{liao2017gapless}, the doped Hubbard model on the square lattice~\cite{corboz2016improved}, and topological orders~\cite{weerda2024fractional, chen2018non, niu2022chiral}. The 2d iPEPS algorithms were extensively applied to SU($N$)-symmetric Heisenberg and Hubbard models on various lattices, including square \cite{bauer2012three, Corboz2011_SU4_dimer, kleijweg2025zigzag, kaneko2024ground}, honeycomb~\cite{corboz2013competing, corboz2012spin, nataf2016plaquette, chung20193}, triangular~\cite{bauer2012three}, and kagome~\cite{corboz2012simplex, xu2023phase} geometries.

The iPEPS tensor networks have recently been further extended to three-dimensional (3d) quantum systems~\cite{vlaar2021simulation, jahromi2019universal, vlaar2023efficient, vlaar2023tensor}. The main difficulty of 3d iPEPS approaches is the calculation of observables. This cannot be done exactly and should be completed using an approximate method. Several approximate solutions were proposed in the literature: First, the Refs.~\cite{jahromi2019universal, jahromi2020thermal, jahromi2021thermodynamics} suggested using simple update mean-field (MF) environments to calculate observables. This method can only be accurate enough for gapped models. Second, the authors of Ref.~\cite{vlaar2021simulation} proposed either employing cluster environments or lower dimensional tensor-network methods, such as 2d boundary iPEPS and the corner transfer matrix renormalization group (CTMRG), to calculate observables. In our recent work~\cite{lukin2024}, we have further generalized the boundary iPEPS + CTMRG approach of Ref.~\cite{vlaar2021simulation} by employing the single-layer mapping for both boundary PEPS and CTMRG~\cite{xie2017, liao2017gapless, lee2018gapless, haghshenas2019single, naumann2025variationally, lan2023reduced}. This allowed us to significantly reduce the computational cost of the observables calculation for 3d iPEPS and reach bond dimensions as high as $D=6$.  Other relevant 3d tensor-network studies include Refs.~\cite{xiang2023, gray2024, latorre2013, braiorr2016phase, king2024computational, xu2025efficient}. 

In this paper, we apply tensor-network methods to the SU($4$)-symmetric Heisenberg model on the cubic lattice. We first introduce the model and discuss the results obtained previously in its 1d and 2d limits. We analyze the highly anisotropic limits and further elaborate on the appearance of previously reported dimerized phases~\cite{Corboz2011_SU4_dimer}. Finally, we analyze the 3d cases, both anisotropic and fully isotropic, using the 3d iPEPS approach developed earlier in Refs.~\cite{vlaar2021simulation,lukin2024}. 
In our previous work~\cite{lukin2024}, we focused on a rather simple 3d lattice problem: the SU(2)-symmetric Heisenberg model on the cubic lattice with a $1\times1\times1$ unit cell. To study the more complex SU(4)-symmetric Heisenberg model, we generalize the proposed method to larger unit cells. We describe the necessary generalizations and apply these to the system under study.

\section{Model}

The main focus of our current study is the SU(4)-symmetric Heisenberg model on the simple cubic lattice. The internal degrees of freedom on the lattice sites (sometimes also called colors) transform in the fundamental representations of the symmetry group (the local Hilbert space on each lattice site has dimension $4$). The Hamiltonian of this model has the following form:
\begin{equation}\label{eq:model}
    H = \sum_{\langle ij \rangle_{\gamma}} J_{\gamma} P_{ij},
\end{equation}
where $\langle ij \rangle_{\gamma}$ denotes summation over the nearest-neighbor pairs of sites along the crystallographic axis $\gamma = \{x, y, z\}$ and $P_{ij}$ is the permutation operator, which acts on the two local basis states on sites $i$ and $j$ as follows: $P_{ij} |\alpha_i \beta_j\rangle = |\beta_i \alpha_j \rangle$. Note that the coupling constants $J_{\gamma}$ can be different; however, if not stated otherwise, we fix $J = J_{x} = J_{y} \neq J_{z}$. The case in which all couplings are equal is referred to as the 3d isotropic point.

In the limit $J_{z} \to 0$, the layers in $xy$ planes decouple  from one another. It is known that in this limit, the ground state of the SU(4) Heisenberg model on the square lattice spontaneously breaks both translational and cyclic rotational symmetry as well as continuous SU(4) spin symmetry. The many-body ground state in this limit is characterized by the $4 \times 2$ unit cell, with different colors forming pairwise dimers \cite{Corboz2011_SU4_dimer}. 

In the opposite limit $J_{z} \to \infty $, the chains of sites in the $z$ direction also decouple from each other and are described separately by the SU(4)-symmetric Heisenberg models on the chain. It is known that the 1d SU(4) Heisenberg chain (completely decoupled from the other neighbors) does not break any symmetries; the system resides in the critical phase, with its low-energy theory described by the Wess-Zumino-Witten (WZW) conformal field theory~\cite{Sutherland1975, Nataf2018}. 

\begin{figure}
    %\centering
    \includegraphics[width=\linewidth]{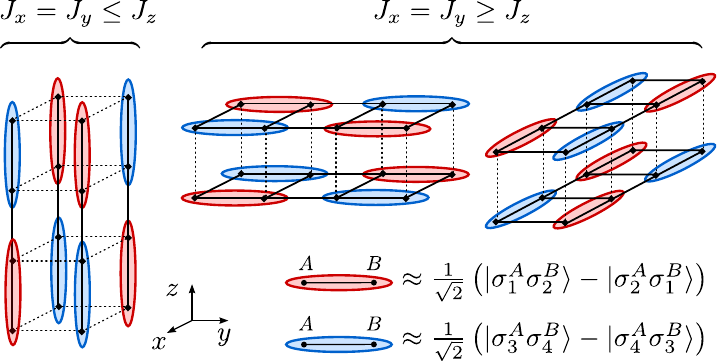}
    \caption{Idealized representation of the ground-state wave function in SU(4)-symmetric Heisenberg model on cubic lattice with spatially anisotropic magnetic couplings. }
    \label{fig:sketch}
\end{figure}

Our main claim in this study is that the dimerized color-ordered phase, which is known from the isotropic 2d limit, persists throughout the entire phase diagram of the model for arbitrary finite $J_{z}/J$ and is continuously connected to the limit of decoupled WZW models, as illustrated in Fig.~\ref{fig:sketch}. We argue that the coupling of SU(4) chains in the transverse directions induces dimerization, which leads to the known ground-state configuration of N\'{e}el-ordered color dimers. 

Let us also briefly comment on the possible physical origin of this configuration using the MF theory argument. For the magnetic product-state configurations, the MF energy expectation value is zero, as upon permutation, these magnetic states are mapped to other magnetic configurations. In contrast, two-color dimers (as depicted in Fig.~\ref{fig:sketch}) are eigenstates of the permutation operators with the eigenvalue $P_{AB}^{(s)}=-1$. This results in substantial energy contribution on dimerized bonds. In turn, the antiferromagnetic arrangement of these dimers yields zero MF energy for all nondimerized bonds: operators on nondimerized bonds permute the colors from one type of dimers to the sites populated by colors from another pair, such that the resulting state overlap is zero. The overall argument for dimerized states cannot be fully convincing since quantum fluctuations can significantly lower the energies of magnetic configurations. However, together with supporting tensor-network numerical analysis confirming stability of the dimerized phase with respect to quantum fluctuations, this argument can serve as explanation.

The dimerized phase can be characterized by a simple order parameter in the form of the difference of energies between dimerized and nondimerized bonds, e.g., in the $z$ direction, which we denote below as $\Delta E_{z}$. 
%%%% FIG. 1 %%%%

\section{Methods}

We employ several tensor-network methods to investigate the model under study. The main reason is that the SU(4)-symmetric Heisenberg models are capable of hosting various highly entangled phases, such as valence-bond solids or spin liquids \cite{Corboz2011_SU4_dimer, corboz2012spin}. To properly describe these strongly correlated regimes, it is necessary to work within a theoretical framework capable of representing highly entangled states. Among such methods, the most widely used are either tensor network or variational Monte Carlo approaches (including the neural quantum state algorithms). In this respect, tensor networks have a slight advantage, as they can be directly applied in the thermodynamic limit. 
Note that, below, we generally employ real-valued tensor networks. For verification purposes, we also developed other alternative optimization schemes with complex-valued tensor networks, which resulted in the same observations. We also do not impose any additional symmetries \change{[both continuous symmetries as SU(4) and discrete symmetries as lattice rotation or reflection]} on our tensor networks. 

Our theoretical analysis consists of three parts. First, we briefly return to the SU(4)-symmetric Heisenberg model on the square lattice, which was already studied in Ref.~\cite{Corboz2011_SU4_dimer}. The difference from the previous study is that here we also consider the anisotropic cases with $J_{x} \neq J_{y}$. We employ 2d iPEPS to obtain results in this regime of anisotropic couplings.  Second, we consider another limit, where $J_{z} \gg J_{x}=J_{y}$ and work in the framework of the decoupled MPS (decoupled chains) ansatz to study this regime with the MPS$\,+\,$MF theory approach~\cite{bouillot2011, bollmark2020, bollmark2023}.  Third, we focus on the genuine 3d isotropic point with $J_{x}=J_{y}=J_{z}$ and employ 3d iPEPS. These three cases are described in detail below. 

\begin{figure}[t]
    \includegraphics[width= \linewidth]{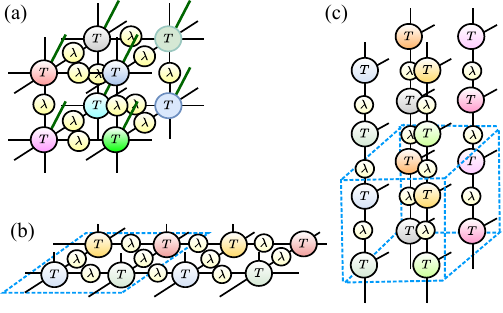}      \caption{\label{fig:tn_ansatzes}%
        Illustration of three different tensor network ans\"atze employed in this study: (a) 3d iPEPS wave function with tensors~$T$ placed on the sites of cubic lattice and bond matrices $\lambda$ obtained naturally within the simple update (note that the matrices $\lambda$ are bond dependent, but the picture is simplified); (b) 2d iPEPS wave function with the $2 \times 2$ unit cell (indicated by surrounding dashed lines); (c) MPS$\,+\,$MF ansatz (with the unit cell $2 \times 2 \times 2$), where the wave function is represented as a product of MPSs formed along the $z$ axis. In all cases the different tensor colors correspond to different tensors inside the unit cell. Note that the unit cells are chosen here for illustrative purposes and may differ from the unit cells used in actual computations. }
\end{figure}

\subsection{2d iPEPS}

We begin our theoretical analysis from the SU(4)-symmetric Heisenberg model on the square lattice with anisotropic couplings ($J_{x} \neq J_{y}$). For this purpose, we employ the standard iPEPS ansatz, as shown in Fig.~\ref{fig:tn_ansatzes}(b). We take the elementary unit cells of size $4 \times 2$ and $4 \times 4$. The iPEPS is initialized with a random product state and then optimized using the simple update method with a gradually decreased time step~\cite{orus2014practical, bruognolo2021beginner, jiang2008}. 
After the optimization, we compute energies for different lattice bonds using the double-layer CTMRG method~\cite{orus2009simulation, nishino1997corner}. 

\subsection{MPS + MF}

We are also interested in the limit where the coupling along one crystallographic axis is much stronger than the couplings in the transverse plane. In this limit, we can view the system as a set of nearly decoupled, weakly interacting chains. 
We expect that the correlations in these chains are strong, as is natural for 1d systems, while the entanglement between different chains should be weak. In the extreme case, we can consider that the interaction between different chains can be included by the MF treatment, while the entanglement along the chain should be fully described with a nonperturbative method. In this limit, we propose using the ansatz consisting of separate MPSs along different chains, as shown in Fig.~\ref{fig:tn_ansatzes}(c). Such a combination of MPS along the chain and MF theory treatment of the interaction between the chains is known as the MPS$\,+\,$MF framework~\cite{bouillot2011, bollmark2020, bollmark2023}. To optimize such an ansatz, we employ infinite time-evolving block decimation (iTEBD), where after the application of imaginary time Trotterized evolution gates between different chains, we truncate the new bond back to the bond dimension $D=1$. Note that, since the chains are decoupled, we can compute all the observables exactly without CTMRG.  During the optimization, we gradually increase the bond dimensions of separate MPS.  

\subsection{3d iPEPS}

In the proximity of the isotropic point of the model in Eq.~\eqref{eq:model}, we cannot expect decoupled chains or 2d iPEPS wave functions to be suitable ans\"atze, as generally, we can expect rather complex 3d correlations. To include such a possibility, we employ 3d iPEPS wave functions, as shown in Fig.~\ref{fig:tn_ansatzes}(a). 
The 3d iPEPS wave function is optimized using imaginary time evolution with the simple update method~\cite{jiang2008, bruognolo2021beginner, orus2014practical}, as described in Refs.~\cite{vlaar2021simulation, jahromi2019universal, jahromi2020thermal, jahromi2021thermodynamics}. \change{As a parameter of convergence, we analyze bond matrices~$\lambda$ on the bonds of the lattice, which naturally appear in the simple update scheme. }

As an initial approximation, we compute observables with the simple update environments (see Ref.~\cite{jahromi2019universal}). This calculation is approximate, but it allows one to estimate the phase of the wave function (perhaps, except for the states with topological order, which is not the case for the SU(4)-symmetric Heisenberg model). In addition, it allows one to pinpoint the necessary unit cell and iPEPS bond dimensions $D$. After this initial step, the observables of iPEPS are computed with a more accurate CTMRG method, which is described below. 

\subsection{Boundary iPEPS and single-layer CTMRG}

\begin{figure}[t]
    \includegraphics[width= \linewidth]{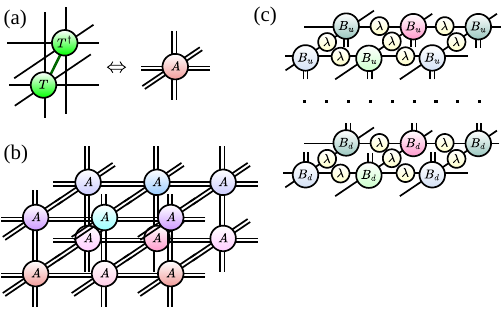}      \caption{\label{fig:3d_contraction}%
        3d double-layer tensor network representing the norm of the 3d iPEPS wave function: (a) From the onsite tensors $T$ we can form the double-layer tensors $A$. Note that additional $\lambda$ matrices on the bonds of 3d iPEPS are assumed to be already absorbed into iPEPS site tensors~$T$; (b) 3d tensor network representing the norm of iPEPS wave function (different colors of $A$ tensors show the different sites in the unit cell, which is chosen as $2 \times 2 \times 2$); (c) the boundary iPEPS approximating the leading eigenvector of the matrix formed by the two horizontal layers of $A$ tensors. We introduce the subscripts $u$ and $d$ in the boundary tensors~$B$ to distinguish between the boundary iPEPS representing the contraction of top and bottom parts of the tensor network, respectively. \change{Note that various bond matrices $\lambda$ are different on different bonds and for different bPEPS, but we show them in a simplified way here. }
         }
\end{figure}

To compute operator expectation values, it is necessary to contract an infinite double-layer 3d tensor network, as shown in Figs.~\ref{fig:3d_contraction}(a) and (b).
%%%%%%%%%%%%%%%%
%%% FIGURE 3 %%%
%%%%%%%%%%%%%%%%
%%%%%%%%%%%%%%%%
%%%%%%%%%%%%%%%%
Such tensor network contraction is performed approximately in two steps. First, the tensor network parts from the top and bottom, which sandwich a chosen horizontal layer, are contracted. Let us first assume that the unit cell has the lowest (i.e., unit) periodicity in the vertical direction. To contract the bottom and top parts of the tensor network, we should note, first, that a horizontal layer of the 3d tensor network can be viewed as a matrix formed by its top to bottom indices. Then, the contraction is equivalent to finding the leading left and right eigenvectors of this matrix. Due to the inherent lattice structure of the matrix, we can assume that the leading eigenvectors can be represented as 2d iPEPS (boundary iPEPS or bPEPS) with bond dimension $\chi_{b}$, as shown in Fig.~\ref{fig:3d_contraction}(c). To find these leading eigenvectors, we employ the power method, as described in detail in our previous work~\cite{lukin2024}. We apply every double layer to the boundary iPEPS in two stages, as discussed in Ref.~\cite{lukin2024}. After each stage, we truncate the bond dimensions of the bPEPS indices back to their original values using superorthogonal canonical form \cite{Ran2012, phien2015b}. 
This transformation into superorthogonal canonical form has a computational cost $D^{7} \chi_{b}^{5}$ and memory scaling $D^{6} \chi_{b}^{4}$. In practice, in our analysis the computational time of the bPEPS part is always smaller than the time required for the CTMRG part, but exactly this bPEPS transformation constitutes the main memory bottleneck. 

We can now weaken our initial assumption of unit periodicity in the $z$-direction and assume that there are $k$ different layers stacked vertically and repeating periodically. In this case, the contraction of the tensor network from the top and bottom is equivalent to finding the leading eigenvectors of the matrix obtained as multiplication of the $k$ layer transfer matrices. To obtain these eigenvectors, we again assume that the eigenvectors can be approximated as 2d iPEPS and then apply these $k$ layer transfer matrices to these 2d iPEPS sequentially (similar to how it is done in the 2d case with the boundary MPS approaches~\cite{Nietner2020efficient}). After application of every layer, we perform truncation of bPEPS bond dimensions. 

In practice, we usually obtain the bPEPS tensors with a gradual increase of $\chi_{b}$. We initialize bPEPS as a product state with $\chi_{b}=1$ and then gradually increase $\chi_{b}=2,4,...$. At every $\chi_{b}$, we check the element-wise convergence of bPEPS bond matrices $\lambda$ up to accuracy $10^{-6}$. It is usually sufficient to perform $5-8$ power-method iterations to converge to this accuracy. In this way we increase $\chi_{b}$ up to values necessary for convergence. In our experience, for observables to converge up to accuracy $10^{-5}-10^{-4}$ the smallest diagonal values of $\lambda$ matrices should be of the order $10^{-5}$. Using these criteria, it is approximately possible to find correct $\chi_{b}$. This $\chi_{b}$ should be additionally checked for convergence using comparison of final results with the previous $\chi_{b}$.

\begin{figure}[t]
    \includegraphics[width= \linewidth]{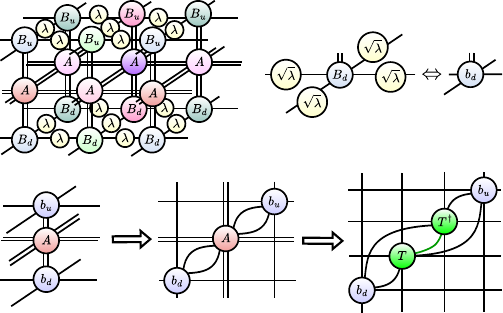}      \caption{\label{fig:sl_ctmrg}%
        Contraction of a residual 2d tensor network consisting of a middle layer with the double-layer tensors~$A$ (see also Fig.~\ref{fig:3d_contraction}) sandwiched between the boundary iPEPS, which represent the top and bottom parts of the tensor network. Different colors of tensors show different points in the projected unit cell, which is chosen here as $2 \times 2$. The resulting four-layer tensor network can be mapped into the effective single-layer tensor network with a 16-times-larger unit cell, as discussed in detail in Refs.~\cite{lukin2024, haghshenas2019single}. 
         }
\end{figure}

After we determined the bPEPS tensor networks, we contract the residual four layer 2d tensor networks shown in Fig.~\ref{fig:sl_ctmrg}. To this end, we first map the resulting tensor network onto the single-layer network, as shown in Fig.~\ref{fig:sl_ctmrg} and discussed in detail in Ref.~\cite{lukin2024}. The main difference in this mapping compared with Ref.~\cite{lukin2024} is that the original unit cell in the $xy$ plane can be nontrivial (e.g., of size $m\times n$ with $m,n\in \mathbb{N}$). Then the mapping results in the single-layer tensor network with the effective unit cell $4m \times 4n$.  This new single-layer tensor network is then contracted with the CTMRG algorithm and auxiliary bond dimension $\chi_{c}$~\cite{nishino1996corner, nishino1997corner, orus2009simulation, corboz2014competing}. Note that one may sometimes encounter the ill-convergence of this CTMRG by starting from random values. To mitigate such potential issues with convergence, we use two different approaches: (i) physical initialization of CTMRG tensors (physical initialization can be set as a product of Kronecker deltas connecting both iPEPS and bPEPS bra and ket indices) followed by a couple of layers of bMPS-based contraction prior to full CTMRG bMPS truncation uses much larger effectively infinite environment for the truncations, which is initially much better than the nonconverged CTMRG environments); (ii) CTMRG with larger environments for the definition of projectors (if for computing projectors one usually contracts $4 \times 2$ blocks formed by the tensor network and CTM tensors~\cite{corboz2013competing}, then in our case, we use $4 \times l$ blocks, where $l>2$). In addition, we usually gradually increase the bond dimension $\chi_{c}$ starting from some moderate values (e.g., $\chi_{c}=30$) up to the maximal values. At every $\chi_{c}$, we perform the fixed number of CTMRG iterations, e.g., seven sweeps per four lattice directions. The relevant observables are then measured at both intermediate and final values of $\chi_{c}$ to check convergence.  

The computational cost of the CTMRG part scales as $\chi_{c}^{3} D^{3} \chi_{b}^{3} + \chi_{c}^{2} (D^{6} + D^{2} \chi_{b}^{4})$. Note that $\chi_{b} \approx 3D$ and $\chi_{c} > D^{2}$. In our previous study~\cite{lukin2024}, we suggested the scaling $D^{12}$ of the whole calculation, but it is difficult to confirm this since it is possible to perform calculations only up to $D=6$ in some cases. In addition to that, the precise scaling of $\chi_{b}$ with $D$ remains not completely clear. For the SU(2) Heisenberg model on the simple cubic lattice the scaling is approximately $\chi_{b} \approx 2D$, but for the SU(4) Heisenberg model it is closer to $\chi_{b} \approx 3 D$ and it may have some nonlinear dependence on $D$. The precise dependencies of $\chi_{c}$ and $\chi_{b}$ on $D$ necessary for convergence will probably be determined only after a sufficient number of models studied with the method. 

In the discussion above, we generally assumed that the tensor network is contracted from the top and bottom of a chosen layer with bPEPS. This contraction allows us to compute operator expectation values for the operators inside this layer. Hence, the CTMRG contraction should be repeated for all $k$ layers. Still, the described construction cannot be used to compute the energy expectation values for operators along vertical $z$ bonds. To deal with such operators, we rotate the tensor network and repeat the calculations above with the rotated network, where the new horizontal layer corresponds, e.g., to the original $xz$ planes. 

Finally, there are additional self-consistency checks, which enable verification of our results. First, the local reduced density matrices are hermitian, but the contraction method described above does not guarantee this hermiticity, since the top and bottom bPEPS are generally different, and the single-layer CTMRG also does not use hermiticity. Hence, the hermiticity of the density matrix is restored only after sufficient convergence of bPEPS and CTMRG---this is one of the nontrivial checks of the validity of the algorithm. Second, as mentioned above, certain observables require additional tensor network rotations to be found, while other observables, such as the onsite reduced density matrices, can be determined by using both the original and rotated tensor network contractions. Therefore, we compute these onsite reduced density matrices by utilizing both mentioned contractions and compare the results to confirm the algorithm convergence and consistency.

\section{Results}

\subsection{2d limit}

We first turn our focus to the systems in the limit of decoupled planes, i.e., square lattices with anisotropic in-plane couplings. In the 2d case, the full machinery of 2d iPEPS is available to us. 

We analyze the 2d limit by taking $J_{z} = 0$ and $J_{x} \neq J_{y}$. Without loss of generality, below we assume that $J_{x} \geq J_{y}$ and $J_{x}=1$. In the isotropic case, it is known that the model is in the color-ordered dimerized phase~\cite{Corboz2011_SU4_dimer}, which spontaneously breaks lattice translational, lattice rotational and continuous spin-rotational SU(4) symmetries. In this subsection, we study the anisotropic case with $J_{y} \in [0,1]$ by applying the iPEPS algorithm and show that the dimerized phase at $J_{x} = J_{y}$ is continuously connected with the limit $J_{y} \to 0$, which is described by the decoupled translationally invariant WZW models, with translational invariance [and SU(4) symmetry] along the chain gradually restored. 

We take the 2d iPEPS wave functions with unit cells of size $4 \times 4$ and bond dimension \change{ $D=10$} as the ground-state ans\"atze. We employ the CTMRG approach with the auxiliary dimension \change{$\chi = 100$} for the computation of two-site correlation functions. 
In all cases, we observe the same structure of the dimerized phase as in the isotropic limit. 

\begin{figure}[t]
    \includegraphics[width= \linewidth]{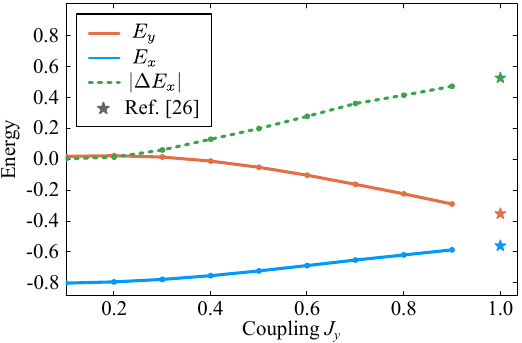}      \caption{\label{fig:2dpeps}%
        Dependencies of the average energies $E_{x}$ and $E_{y}$ on $x$ and $y$ bonds, respectively, and the average difference of energies between the two consecutive $x$ bonds $ |\Delta E_{x}|$ on the strength of interaction on the $y$ bonds $J_{y}$ in the 2d limit of SU(4) Heisenberg model (with $J_{x}=1$). The calculations are performed within the 2d iPEPS with the bond dimensions \change{$D=10$}. \change{For comparison, we also show the results at $J_{x}=J_{y}=1$ and $D=10$ from Ref.~\cite{Corboz2011_SU4_dimer} (see Fig.~3 therein) as star symbols}.  }
\end{figure}

In Fig.~\ref{fig:2dpeps}, we show the energy expectation values along the $x$ direction (along the dimer orientation) for both strongly and weakly correlated bonds. Note that, here and below, the bond-specific energies $E_{ij,\gamma}=\langle H_{ij,\gamma}\rangle$ are additionally averaged over all similar bonds in the given unit cell. We can conclude that the breaking of translational symmetry, as well as the continuous SU(4) symmetry, is a consequence of the addition of a (small) coupling between different chains. Moreover, the greater the interchain coupling~$J_y$, the more pronounced the breaking of translational symmetry along the chain becomes. It means that the coupling between different chains is the reason for the dimerized color-ordered phase, which remains stable including the 2d isotropic point.

\subsection{1d limit}

We proceed to study the limit \change{$J_{x} = J_{y}=J \to 0$} while keeping the coupling between chains along the $z$ direction fixed ($J_z=1$). The decoupled chains are described by the WZW models and are translationally invariant. As discussed in the previous subsection, the addition of a small coupling between the chains may naturally lead to dimerization. In the MF description, the interaction with the four nearest-neighbor chains is simply the same as the interaction with the two chains with the twice larger couplings. Hence, we can expect that the behavior of coupled chains can be qualitatively similar on both highly anisotropic square and cubic lattices (with the difference that the cubic lattice may potentially have more complex unit cells).  

\begin{figure}[t]
    \includegraphics[width= \linewidth]{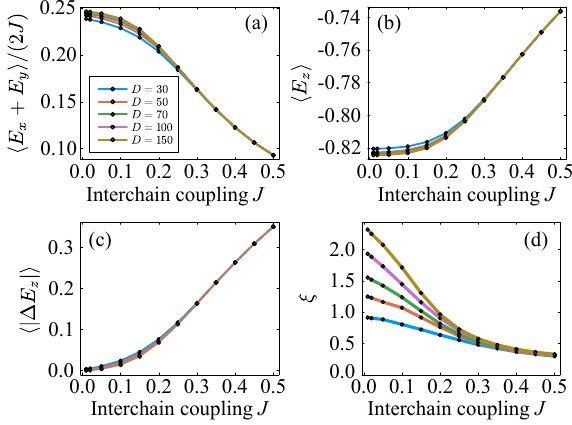}      \caption{\label{fig:3dchains}%
        Dependencies of \change{observables: (a)} The average energy on the $x$ and $y$ bonds $E_{x,y}/J$ \change{and (b)} on the $z$ bonds $E_{z}$ {(in units of $J_{z}=1$)}, \change{(c)} the average difference of energies between the two consecutive $z$-bonds $ |\Delta E_{z}|$\change{, and (d) the correlation length~$\xi$} on the coupling in transversal directions $J\equiv J_{x,y}$ in the 3d chain regime ($J<J_z$) of the SU(4)-symmetric Heisenberg model. The calculations are performed within the 3d chain MPS \change{with different bond dimensions up to  $D_z=150$.} }
\end{figure}

We study the behavior in this limit within the decoupled chain MPS + MF ansatz. Our results for the energy dependence on different lattice bonds are shown in Fig.~\ref{fig:3dchains} on the cubic lattice with unit cell $4 \times 4 \times 4$. It is clear that, in both cases, dimerization develops as a consequence of interchain coupling. The results are obtained for the MPS bond dimensions up to $D_z=150$. For moderate interchain couplings $J \approx 0.25-0.5$, the results are converged in bond dimension. For very small $J$, the local operators are almost converged, but the correlation length~$\xi$ shown in Fig.~\ref{fig:3dchains}(d) continues to increase. This behavior is expected, as in the limit $J \to 0$ the correlation length should diverge. Note that the  employed method in this study may be not optimal in the regime of very small couplings and nearly critical system; it should become more efficient if one extensively uses symmetries as SU(4) and probably applies some version of DMRG instead of iTEBD. 

We would like to comment slightly more on the correlation length behavior. In principle, dimerization spontaneously breaks the SU(4) symmetry and the full system correlation length should diverge for every value of interchain coupling~$J$. It seems that the coupled MPS ansatz is incapable of describing this behavior with sufficient accuracy, and for this reason the correlation length stabilizes at finite values at higher interchain couplings.

Note that we cannot strongly confirm that dimerization develops immediately after addition of nonzero interchain coupling. At very small interchain coupling, our iTEBD-based calculations are slightly noisy due to the large size of the unit cell and small couplings between chains. This does not allow us to prove that dimerization develops for any interchain coupling and not at some nonzero interchain coupling $J \approx 0.1$.

We should also emphasize several details here. First, at small interchain coupling ($J_{x,y}\lesssim0.1J_z$), the energy per bond smoothly connects to the WZW model with a uniform distribution of energies along the chain with $E_z\approx-0.825$~\cite{Nataf2018}. Second, the expectation values of the exchange operators $P_{x,y}=E_{x,y}/J$ between different chains (which are computed in a MF-like fashion, since the neighboring chains are not entangled) decrease from $0.25$ to $0.1$. The limit with $P=0.25$ corresponds to the probability that the sites in two different chains have the same color when the colors are distributed homogeneously along the chain. The decrease of this expectation value to $0.1$, even at the level of MF theory, points to dimerization in the chains. Finally, we observe that for the sites on the same chain, the expectation value of the operator $P$ differs between even and odd bonds. This also points toward the dimerization of the chains.

\subsection{Isotropic point in 3d}

From the two previous subsections, it is clear that dimerized color-ordered states appear in the cubic lattice both in the limit of large $J_{z}$ and in the opposite limit $J_{z} \to 0$. The natural question is whether these two regimes are adiabatically connected to each other through the isotropic point. In this subsection, we present the evidence that this is, indeed, the case. For this purpose, we study the system at the isotropic point within the 3d iPEPS approach. 

We first optimize the iPEPS wave function with the simple update on the $4 \times 4 \times 4$ lattice and with bond dimensions up to $D=8$. The choice of the unit cell with all dimensions being multiples of $4$ is natural for the model with the SU(4) symmetry group since it can host the relevant dimerized or magnetic orderings, as those discussed in Ref.~\cite{Unukovych2024}. To approximately pinpoint the phase, we calculate the observables using the simple update environments. 
Note that it is possible to complete simple update calculation even for larger bond dimensions (up to $D=12$), but it is currently too costly to compute their expectation values with controlled methods. One can still obtain some reliable information on the state from the converged simple update tensors. For example, it looks realistic to compute observables based on approximate simple update environments \cite{jahromi2019universal, jahromi2021thermodynamics} or to inspect bond spectra in the superorthogonal canonical form \cite{braiorr2016phase}. Both the approximate observables and bond spectra allow one to approximately infer the nature of the phase but do not give systematically controlled results. 

Our numerical analysis suggests that starting from $D=4$, the wave function converges to the dimerized phase with the $2 \times 2 \times 4$ unit cell. Note that we use different initialization with random product states to check that simple update converges to the same phase. After this preliminary estimation of the necessary unit cells and bond dimensions, we focus on the $2\times2\times4$ unit cell with the bond dimensions $D=4$ and $5$. The maximal bond dimension $D=5$ was chosen 
because, for bond dimensions $D \geq 6$, the necessary computational cost for the tensor-network contraction with CTMRG becomes too large. The results below are shown for the $D=5$ and $4$ wave functions. Note that these results may be criticized from different points of view. The first critical point is that the local Hilbert space dimension is 4, and bond dimension $D=5$ is only moderately larger. The second important point is that the simple update may be insufficient to detect some phases. In our defense, we could say that the simple update approach was also employed in the 2d SU(4)-symmetric Heisenberg model in Ref.~\cite{Corboz2011_SU4_dimer}. 

Next, we employ the boundary iPEPS + CTMRG approach to calculate observables of the optimized 3d iPEPS wave function. We calculate the local onsite reduced density matrices for all sites inside the unit cell as well as the energies on the bonds. It is necessary to converge boundary iPEPS in $z$ and $y$ directions to calculate energies on all bonds. 
For the $z$ direction, the maximal boundary iPEPS bond dimension for the $D=5$ wave function is fixed to $\chi_{b} = 14$, while for the $y$ direction, the maximal boundary iPEPS bond dimension is fixed to $\chi_{b}=17$. For the $D=4$ wave function, the maximal $\chi_{b}$ values in $z$ and $y$ directions are, respectively, $10$ and $14$.
For these bond dimensions, the smallest boundary iPEPS bond spectra $\lambda$ are of the order $10^{-5}$, which is used as a criterion of bPEPS convergence. For converged bPEPS, we finally calculate the observables using the single-layer CTMRG, with the CTMRG bond dimensions up to $\chi_{c}=160$ \change{for $D=5$ and up to $\chi_{c}=120$ for $D=4$}. The corresponding bond energies converge reaching accuracy in the range $10^{-5} - 10^{-4}$, while the local reduced density matrix elements converge with accuracy below $10^{-5}$.
Note that the local reduced density matrices also converge with respect to hermiticity up to errors of the order of $10^{-6}$, and the difference between local reduced density matrix elements computed with the original and rotated bPEPS is of the order of $10^{-5}$. 

We observe that the bonds can be divided into three groups according to their respective energies: The first group consists of $x$ and $y$ bonds; the second group includes energies on $z$ bonds starting from even $z$ coordinates; the third group includes the rest of the $z$ bonds. The average energy of bonds in the first group is equal to $ E_{x} \approx E_{y} = -0.1823$ (for $D=4$ it is equal to $E_{x} \approx E_{y} = -0.1704$). The energies of bonds in the second group have the average energy $ E_{z, {\rm even}} = -0.8668$ (for $D=4$ $E_{z, {\rm even}} = -0.8003$), while the energies in the third group are equal to $ E_{z, {\rm odd}} = - 0.1908$ (for $D=4$ $E_{z, {\rm odd}} = -0.1779$). Note that, as various rotational and translational symmetries are not enforced during the optimization and simple update iPEPS optimization being not very accurate, there were some residual discrepancies between energies on bonds of the same group, especially for the bonds in the first and third groups.  

\begin{figure}[t]
    \includegraphics[width= 0.9\linewidth]{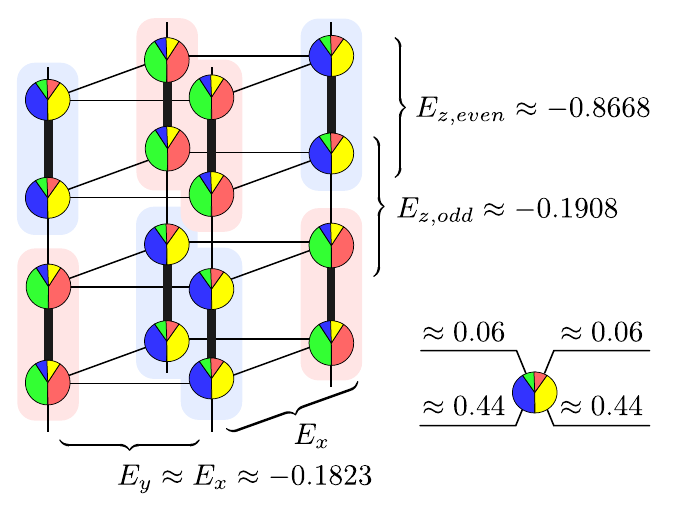}      \caption{\label{fig:colors}%
        Sketch of spin-flavor concentrations in the $2 \times 2 \times 4$ unit cell of the isotropic cubic lattice. Bold lines correspond to the bonds with the lowest energy $E_{z, {\rm even}}$. The results are obtained by 3d iPEPS with the bond dimension $D=5$. }
\end{figure}
In Fig.~\ref{fig:colors}, we show the onsite concentrations at $D=5$ of different colors on different sites of the unit cell and the bonds with the largest energies (these we call above bonds of the second kind). The concentrations of colors on the site approximately follow the proportions $0.447:0.442:0.057:0.054$. The slight difference between the first and second concentrations is likely due to the limitations of the simple update at the bond dimension $D=5$. Still, we can conclude that the concentrations of the first two and the last two colors are approximately equal on all sites. In Fig.~\ref{fig:colors}, we can clearly see that the colors form either red-green or yellow-blue dimers with  large negative energies.

\section{Conclusions and Outlook}

We studied many-body ground states of the SU($4$)-symmetric Heisenberg model on a simple cubic lattice using different tensor network approaches by employing 3d iPEPS variational wave functions. For this wave function ansatz, we extended our previous method of calculating observables to handle large unit cells. Our numerical results point toward the persistence of the dimerized color-ordered phase known from the 2d limit~\cite{Corboz2011_SU4_dimer} in both anisotropic and completely isotropic 3d regimes. This, in turn, is a sign that the 3d SU($N$)-symmetric lattice systems remain very promising as hosts for exotic quantum phases beyond the MF theory results~\cite{Unukovych2024}. 

There are several possible additional research directions. One of them is the extension of the method to different lattice geometries. The simplest lattices to extend our calculations are hyperhoneycomb and diamond lattices, as the tensor networks on these lattices can be easily mapped onto the simple cubic lattice after the combination of several tensors into one. It is also possible to develop algorithmic extensions to pyrochlore and octahedron lattice geometries (3d Lieb lattice) after switching from iPEPS to the infinite projected simplex states (iPESS) ansatz~\cite{Xie2014}. Some other cases, such as the face centered cubic lattice, may also be within reach after the inclusion of longer range interactions. In this respect, the most straightforward direction to proceed is to study the Heisenberg model on the pyrochlore lattice~\cite{astrakhantsev2021} and some additional models on the hyperhoneycomb lattice. Note that it was previously proposed that some SU($4$)-symmetric models may be realized on the hyperhoneycomb lattices \cite{Natori2018}. 

In addition to potential extensions to other lattice geometries, it is also possible to further develop the algorithm on the cubic lattice. One potential extension is the implementation of the full update optimization scheme \cite{Jordan2008, phien2015} for the 3d iPEPS wave functions. The cost of such optimization is higher but technically manageable. In conditions where we are restricted to small bond dimensions, the energy benefit of the full update scheme can become substantial. A lighter version of this approach is the implementation of a 3d extension of the neighborhood tensor update from Ref.~\cite{Dziarmaga2021}, which combines higher accuracy than the simple update with a still reasonable cost even in 3d. Another possible algorithmic development concerns the calculation of operator averages. In this study, we employed only the simple update environments to optimize bPEPS tensors; however, it is also possible to use the full update. It may also be interesting to compare our observable calculations with the loop expansion modification of belief propagation from Ref.~\cite{evenbly2024loop}. Furthermore, one can also investigate issues with potential CTMRG ill-convergence and compare the observable calculations from the rotated tensor networks to analyze algorithmic convergence.   

\acknowledgments

The authors acknowledge support from the National Research Foundation of Ukraine under the call ``Excellent science in Ukraine 2024–2026'', project No.~2023.03/0073. 

\bibliography{refs}

\end{document}